\documentclass[aps,prd,preprint,nofootinbib]{revtex4}
\usepackage{amsmath,amssymb,graphics,epsfig,multirow}
\usepackage{verbatim}
\usepackage{slashed}

\newcommand{\newc}{\newcommand}
\newc{\beq}{\begin{equation}}
\newc{\eeq}{\end{equation}}
\newc{\eeqa}{\end{eqnarray}}
\newc{\beqa}{\begin{eqnarray}}
\newc{\bsym}{\boldsymbol}
\newc{\mrm}{\mathrm}
\newc{\ovl}{\overline}
\newc{\ovla}{\overleftarrow}
\newc{\ovra}{\overrightarrow}
\newc{\ra}{\rightarrow}
\newc{\lra}{\leftrightarrow}
\newc{\wtil}{\widetilde}
\newc{\eps}{\epsilon}
\newc{\hc}{\dagger}
\newc{\pd}{\partial}
\newc{\SL}{\!\!\!/}
\newc{\LH}{\hat{L}}
\newc{\RH}{\hat{R}}
\newc{\sWsq}{\sin^2\theta_\mathrm{W}}
\newc{\cWsq}{\cos^2\theta_\mathrm{W}}
\newc{\half}{\frac{1}{2}}
\newc{\hh}{\hat{H}}
\newc{\hphi}{\hat{\Phi}}
\newc{\nonr}{\nonumber}

\DeclareMathOperator{\diag}{diag}

\begin{document}
\title{Neutrino Masses via  the Zee Mechanism  in 5D split fermions model}

\author{We-Fu Chang}
\email{wfchang@phys.nthu.edu.tw}
\author{ I-Ting Chen}
\author{Siao-Cing Liou}

\affiliation{Department of Physics, National Tsing Hua University,
HsinChu 300, Taiwan}
\date{\today}

\begin{abstract}
We study the Zee model in the framework of the split fermion model in
$M_4\times S_1/Z_2$ spacetime.
Neutrino masses are generated through 1-loop diagrams without the right-handed neutrinos introduced.
By assuming an order one anarchical complex 5D Yukawa couplings, all the
effective 4D Yukawa couplings
are  determined by the wave function overlap between the split fermions
and the bulk scalars in the fifth dimension.
The predictability of the Yukawa couplings is in sharp contrast to the original Zee model in 4D where
the Yukawa couplings are unknown free parameters.
This setup exhibits a geometrical alternative to the lepton flavor symmetry.
By giving four explicit sets  of the split fermion locations, we demonstrate
that it is possible to simultaneously fit the
lepton masses and neutrino oscillation data  by just a handful free parameters without much fine tuning.
Moreover, we are able to
make definite predictions for the mixing angle $\theta_{13}$,
the  absolute neutrino masses, and the lepton flavor
violation processes for each configuration.
\end{abstract}

\maketitle
\section{Introduction}
In the Standard Model(SM), neutrinos are strictly massless due to the accidental  global $U(1)_{B-L}$ symmetry.
Therefore, to explain the observed nonzero active neutrino masses, one needs to go
beyond  the SM  to introduce  new lepton number violating interaction.

One famous mechanism to generate neutrino mass  was proposed by Zee\cite{Zee:1980ai} thirty years ago.
In addition to the SM Higgs doublet $\Phi_1$,
Zee simply extended the scalar sector  to include  one more  $SU(2)_L$ Higgs doublet $\Phi_2$ and an extra
$SU(2)_L$ singlet charged Higgs $h$. There is no known principle to forbid
the charged singlet Higgs couples to the $SU(2)$ singlet formed by a pair of  lepton doublets.
The appropriate terms in the Lagrangian are
\beq
{\cal L}_{Zee} = -f^1_{ab}\bar{\Psi}_{aL} \Phi_1 e_{bR}-f^2_{ab}\bar{\Psi}_{aL} \Phi_2 e_{bR}
- M_{12} \Phi_1 i\tau_2 \Phi_2 h^* -f^h_{ab} \overline{\Psi^{\textbf{c}}_{aL}} i\tau_2 \Psi_{bL} h + H.c.\,,
\eeq
where $a,b$ are the generation indices, and all the rest are in standard notation.
The Higgs potential is omitted here because it is irrelevant to our present discussion.
This lepton number violating coupling
term $\overline{\Psi^c} \Psi h$ is the key  for generating the effective neutrino Majorana masses.

Since the neutrinos do not receive tree level masses in this setup, the neutrino masses
must be generated by radiative corrections. Hence, the Zee model provides a natural
and economical explanation of the smallness of neutrino masses.
However, the flavor changing neutral current(FCNC) is a well known phenomenological problem in  models with two Higgs doublets.
This consideration led Wolfenstein  to further assume that  the coupling between the leptons and
the Higgs doublets conserves flavor by imposing a discrete symmetry that allows only  $\Phi_1$ to couple to the lepton, namely, $f^2_{ab}=0$ \cite{Wolfenstein:1980sy}.
This simplified version is the so-called  Zee-Wolfenstein model, and it has been studied extensively in the past
thirty years\cite{WZeePheno}. In addition, it has been shown that the Zee-Wolfenstein model is  inconsistent with the accumulated neutrino
experimental data. But it doesn't mean the original version of the Zee model is ruled out. It was found that
it is still possible to accommodate the observed neutrino data when both Higgs doublets are allowed to couple
to the leptons\cite{FullZee}.
However, all these attempts suffer from having too many arbitrary parameters
\footnote{There are $21$ unknown complex Yukawa coupling constants in the original Zee model
for three generations of fermions. } and the fine tuning problem to avoid the
persist FCNC.

On the other hand, the possibility that there exists extra  spacial dimensions has opened up
an  intriguing avenue to connect the fermion flavor problem with the fermion wave function profile in the extra dimension(s).
The hierarchical Yukawa couplings in 4D theories can be resulted from
different level of the  wave function
 overlap between the left-handed and the right-handed chiral fermions in extra dimension ( for recent
 studies along this line in the Randall-Sundrum model see \cite{RSflavor} ).
For simplicity, we choose the split fermion model\cite{SF} to illustrate this idea.
In the following, we shall show that by embedding the original Zee model, without any right-handed neutrinos
\footnote{There are various setups to incorporate the right-handed neutrinos into the split fermions framework
to explain the neutrino masses and mixings, see\cite{RHNuSF}.}, into a 5D  split fermion model,
we are able to accommodate both the charged lepton and neutrino  masses and the observed bi-large mixing angles, and
a naturally suppressed  FCNC as well with a handful of parameters and some  simple orbifolding arrangements. We note by passing that the same
method can be applied to any extra dimension model with nontrivial bulk fermion profiles.

\section{Model Setup}
We now proceed to detail  the model setup. The space-time of the model is described by an $M_4\times S_1/Z_2$
orbifold and   the fifth dimensional coordinate is denoted as $y$.
The physical region is defined as $0\leq y \leq \pi R$, where $R$ is the radius of the compactified
extra spatial dimension.
In the model, all the SM fields are propagating in the bulk whereas the fermions are localized in the fifth
dimension with an unspecified potential.
The orbifolding $Z_2$ transforms $y \leftrightarrow -y$, hence each bulk
degree of freedom  must be  either even or odd  under this $Z_2$.  The
first four (fifth) components of the SM gauge fields must be $Z_2$-even(odd) in order to
reproduce the SM in low energy, and the bulk SM gauge fields
can be spanned by either $\cos(n  y/R)$ or $\sin(n  y/R)$ with the proper normalization.
For simplicity,  we assume  that all the SM chiral fermions are Gaussian distributed  in
$y$, and  the width $1/\mu \equiv \sigma$ is universal.
Assuming that $\sigma \ll R$ ( we set $\sigma/R= 5\times 10^{-4}$ for our numerical study),
the Gaussian distribution can be normalized as
\beq
{\bf g}(y,c^{L/R}_i)= \left( {2\mu^2 \over \pi} \right)^{1/4}  \exp \left[-\mu^2 (y-c^{L/R}_i)^2  \right]\,,
\eeq
where $c^{L/R}_i$ is the location where the chiral split fermion peaks in the fifth dimension.
Since
\beq
{\bf g}(y,c_i) {\bf g}(y, c_j) =\exp\left[ -\frac{\mu^2}{2}(c_i-c_j)^2\right] {\bf g}^2\left(y, \frac{c_i+c_j}{2}\right),
\eeq
the hierarchical 4D Yukawa couplings is attributed to the  Gaussian overlap between two chiral fermions.
The overlap is exponentially suppressed as their relative distance $\triangle c$.
The further apart the two chiral fermions are the smaller their 4D Yukawa.

Since we are mainly interested in the neutrino masses,  the quark sector will be left out of our discussion.
To distinguish from the 4D field, the bulk field will be denoted with a hat.
As advertised, the 5D SM lepton doublet $\hat{\Psi}_{aL}$ and  lepton singlet $\hat{e}_{aR}$ can be expressed as
the product of the 4D wave function and the Gaussian profile in fifth dimension: $\hat{\Psi}_{aL}(x^\mu, y) =\Psi_{aL}(x^\mu) {\bf g}(y, c^L_a)$
and $\hat{e}_{aR}(x^\mu, y) =e_{aR}(x^\mu) {\bf g}(y, c^R_a)$, where $a$ is the generation index.
We also assume that the split fermions reside in a fat brane erected  at the orbifold fixed point $y=0$.
 Since $\sigma\ll R$, their Kaluza-Klein(KK)
modes are much heavier, $\sim 1/\sigma$, than those of the gauge  and scalar bosons level by level and we shall not
include them here.

To enable the Zee mechanism, there are two doublet scalar $\hat{\Phi}_1$, $\hat{\Phi}_2$,
and a Zee singlet $\hat{h}$ propagating in the extra dimension as well.
We assign the bulk scalar fields $\{\hat{\Phi}_1,\hat{\Phi}_2,\hat{h}\}$ to be $Z_2$-(even, odd,odd).
 The bulk scalar fields  $\hat{\Phi}_2, \hat{h}$ can be KK expanded
in terms of $\sin(n y/R)$. Similarly, $\hat{\Phi}_1$ can be KK expanded by
$\cos(n  y/R)$. The $n-$th KK mode has mass $m_{n}= \sqrt{\frac{n^2}{R^2} +m_0^2}$ for both parities, where $m_0$ is
the corresponding  bulk mass parameter in the 5D Lagrangian.
By choosing the parity assignment,  only $\hat{\Phi}_1$ has the zero mode, $\Phi^{(0)}_1$.
And only  $\Phi^{(0)}_1$ can develop a nonzero vacuum expectation value (VEV), $\langle \Phi_1^{(0)} \rangle =v/\sqrt{2}$,  $v \simeq 250$ GeV.
The $\Phi^{(0)}_1$  is identified as the SM Higgs. Its mass $m_{\Phi_1,0}$ is unknown but believed to be less than
or close to the electroweak scale $v$. Therefore, it is tempting to expect the remaining  two
 bulk masses $m_{\Phi_2,0}$ and $m_{h,0}$
are also in the same mass range as $m_{\Phi_1,0}$.   As for the other  physical Higgs,
their masses are all above $1/R \sim {\cal O}($ TeV ).
Since only $\Phi^{(0)}_1$ can contribute to fermion masses,
 the fermion Yukawa couplings to $\Phi^{(0)}_1$ is automatically diagonal after the mass diagonalization.
Then the dangerous FCNC will  be only mediated by the KK excitations hence being suppressed compared to
the usual two Higgs Doublets Model (2HDM) in 4D.

A more subtle reason of choosing this parity assignment
is that  the  profiles of $\Phi_2$ and $h$, $ \sin (n y/R) $,  fall off near $y=0$,
which naturally results in a  smaller 4D Yukawa $f^2$ and $f^h$. With an order one 5D Yukawa couplings,
this miraculously makes neutrino masses roughly  the desired  order of magnitude
 without fine turning.

The relevant Lagrangian for 5D Zee model is given by
\beqa
{\cal L}_{5DZee}&=& -\sqrt{2\pi R} \hat{f}^1_{ab} \overline{\hat{\Psi}_{aL}}
\hat{\Phi}_1\hat{e}_{bR}
-\sqrt{2\pi R} \hat{f}^2_{ab} \overline{\hat{\Psi}_{aL}}
\hat{\Phi}_2\hat{e}_{bR}\nonr\\
&& -\sqrt{2\pi R} \hat{f}^h_{ab} \overline{\hat{\Psi}^{\rm c}_{aL} } i\tau_2
\hat{\Psi}_{bL }\hat{h}
-{\kappa \over \sqrt{2\pi R}} \hat{\Phi}_1 i\tau_2 \hat{\Phi}_2 \hat{h}^* +H.c.\,,
\eeqa
where the  factor $\sqrt{2\pi R}$, of mass dimension $[-1/2]$, is used to make the coupling constants
$\hat{f}$'s and $\kappa$
 dimensionless.
After integrating over $y$ and assuming that  ${\bf g}^2(y,c)\sim \delta(y-c)$  for $\sigma \ll R$, we obtain the effective 4D Yukawa in the Zee model as:
\beqa
f^{1(n)}_{ab}&\simeq& (\sqrt{2})^{1-\delta_{n,0}}\,\hat{f}^1_{ab} \exp\left[ {-(c_a^L-c_b^R)^2 \over 2\sigma^2} \right]\cos\frac{n(c^L_a+c^R_b)}{2R}\,,\nonr\\
f^{2(n)}_{ab}&\simeq& \sqrt{2}\,\hat{f}^2_{ab} \exp\left[ {-(c_a^L-c_b^R)^2 \over 2\sigma^2} \right]\sin\frac{n(c^L_a+c^R_b)}{2R}\,,\nonr\\
f^{h(n)}_{ab}&\simeq& \sqrt{2}\,\hat{f}^h_{ab} \exp\left[ {-(c_a^L-c_b^L)^2 \over 2\sigma^2} \right]\sin\frac{n(c^L_a+c^L_b)}{2R}\,,
\label{eq:4DYukawa}
\eeqa
where ``$(n)$'' in the superscript labels the KK level.
Furthermore, there is a tower of  cubic  terms for the KK Higgs in the effective 4D Lagrangian,
\beq
{\cal L}^{New}_{Zee} \supset
- {\kappa\over 2\pi R} \sum_{n,m=1}^\infty \delta_{n,m}\, \Phi_1^{(0)} i\tau_2 \Phi_2^{(n)} h^{(m)*} + H.c.\,,
\label{eq:4Dcubic}
\eeq
where the Kronecker delta  is due to the orthonormality of eigenmode $\sin(ny/R)$.
The coupling $\kappa$ plays the same role as $M_{12}$ in the original Zee model and  it controls the overall size of neutrino masses as well.

After electroweak symmetry breaking, $\langle \Phi_1^{(0)} \rangle =v/\sqrt{2}$,
 the charged lepton mass matrix element can be read as
\beq
{\cal M}^e_{ab} = \hat{f}^1_{ab} \frac{  v }{\sqrt{2}}\exp\left[ {-(c_a^L-c_b^R)^2 \over 2\sigma^2} \right]\,.
\eeq
Assuming that all 5D Yukawa couplings, $\hat{f}^1_{ab}$, are of the same order,
the charged lepton mass hierarchy becomes a problem of finding the solution of
the split fermion peak locations $\{ c^R_{1},c^R_{2}, c^R_{3},c^L_{1},c^L_{2},c^L_{3}  \}$ in the fifth dimension.
The mass matrix  can be brought diagonal by a bi-unitary transformation,
\beq
\diag\{m_e, m_\mu, m_\tau\} =U_L^\dag {\cal M}^e U_R\,.
\eeq
In the charged lepton mass basis, the Yukawa couplings become $f^{2(n)} \Rightarrow U_L^\dag f^{2(n)}U_R $, and
 $f^{h(n)} \Rightarrow U_L^T f^{h (n)}U_L $.

\begin{figure}[htb]
\centering
\includegraphics[width=4in]{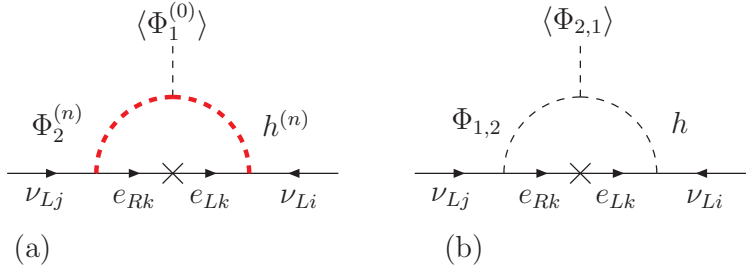}
\caption{ Comparison of 1-loop diagrams  for giving neutrino masses in (a) 5D split fermion Model,
and (b) in the 4D original Zee Model.
Red color (thick dashed line) is used for the KK excitations. \label{fig:1loop} }
\end{figure}

Neutrino Majorana masses can be generated by an 1-loop diagram very similar to the case in the original 4D Zee model,
see Fig. \ref{fig:1loop}-(a).
From Eq.(\ref{eq:4Dcubic}), it is clear that  only  $\Phi^{(n)}_2$ and $h^{(n)}$ of
the same KK level can run in the loop.
In comparison with the original 4D Zee model where both $\langle \Phi_1\rangle$ and
$\langle\Phi_2\rangle$ contribute, see Fig. \ref{fig:1loop}-(b),
only the VEV of $\Phi_1^{(0)}$ contributes in the 5D version.
Another difference is that we need to  sum up  the contributions from all KK excitations in the 5D Zee model.
In practice, the KK summation shall be cut off at some large number
before the effective theory treatment becomes invalid.
We take the cutoff $\sim 0.1 (1/\sigma) \sim 200 /R$ to be consistent with the approximation
that the KK split fermions are ignored.
However, the result is not very sensitive to the choice
due to the $1/n^2$ dependence.\footnote{In fact, the infinite sum converges,
$\sum_{n=1}^\infty \cos(n\pi \triangle)/n^2 = \frac{\pi^2}{6}(1-3|\triangle|+3 \triangle^2/2)$, for $|\triangle|\leq 2$.}

  The mixing between $\Phi^{(n)}_2$ and $h^{(n)}$ is about ${\cal O}(v R/n^2)$ of  their mass square, $\sim(n/R)^2$, thus
it can be ignored in calculating the loop diagram, Fig. \ref{fig:1loop}-(a).  In the charged lepton mass basis,
the resulting effective neutrino mass matrix element can be found to be:
\beq
{\cal M}^\nu_{i j} \sim \frac{1}{16\pi^2}\sum_{n=1}^{\infty} \sum_{k=e,\mu,\tau}
\left(\frac{v}{\sqrt{2}}\right) \left(\frac{\kappa}{2\pi R}\right){ m_k   \left(f^{2(n)}_{ik}\right)^* f^{h(n)}_{kj}  \over M^2_{\Phi_2,n} -M^2_{h,n}} \ln \frac{M^2_{\Phi_2,n}}{M^2_{h,n}}+ \left( i \leftrightarrow j\right)\,.
\label{eq:ZeeM1}
\eeq
For $1/R \gg M_{\Phi_2,0}, M_{h,0}$, the dependence of $M_{\Phi_2,0}, M_{h,0}$ disappears and Eq.(\ref{eq:ZeeM1}) takes a simpler form
\beq
{\cal M}^\nu_{i j} \simeq \frac{1}{16\pi^2}\sum_{n=1}^{\infty} \sum_{k=e,\mu,\tau}
\left(  { \kappa v R \over 2 \sqrt{2}\pi } \right) \frac{m_k}{n^2} \left[ \left(f^{2(n)}_{ik}\right)^* f^{h(n)}_{kj} + \left(f^{2(n)}_{jk}\right)^* f^{h(n)}_{ki} \right]\,.
\label{eq:ZeeMsimp}
\eeq
In this large $1/R$ limit, the neutrino mass is NOT sensitive to the bulk bare masses of $\hat{\Phi}_2$ and $\hat{h}$.

Once the neutrino mass matrix being diagonalized by the Pontecorvo-Maki-Nakagawa-Sakata (PMNS) matrix $U_\nu$,
\beq
\diag\{m_1, m_2, m_3\} = U_\nu^\dag {\cal M}^\nu U_\nu^*\,,
\eeq
the three neutrino mixing angles and the 3 CP violating phases  can be extracted from  $ U_\nu$.

\section{Numerical Result}
Combining all neutrino data, the current best fit results at $99.73\%$ CL are \cite{PDG}
\beqa
2.07\times 10^{-3}\; \mbox{eV}^2 \leq |\triangle m_{31}^2| \leq 2.75\times  10^{-3}\; \mbox{eV}^2\,,\nonr\\
7.05\times 10^{-5}\; \mbox{eV}^2 \leq \triangle m_{21}^2 \leq 8.4\times  10^{-5}\; \mbox{eV}^2\,,\nonr\\
0.36 \leq \sin^2 \theta_{23}  \leq 0.67 \,,\;
0.25 \leq \sin^2 \theta_{12}  \leq 0.37\,,\;
\theta_{13}  \leq 0.23\; \mbox{rad}\,,
\eeqa
or equivalently $\sin^2 2\theta_{23}\geq 0.88$ and $ 0.75 \leq \sin^2 2\theta_{12}\leq 0.93$.
The goal of numerical is to fit the charged lepton masses and all
neutrino data  simultaneously.
As an illustration, we use $1/R = 1 $ TeV,  $\sigma/R =5\times 10^{-4}$, and
 the KK summation is taken for the first 200 excitations as explained earlier.
We take  $M_{h,0} = 400$ GeV  and $M_{\Phi_2,0} = 200$ GeV as the input. However we stress again
that our result is insensitive to the choice of $M_{h,0}$ and $M_{\Phi_2,0}$,
 as shown in Eq.(\ref{eq:ZeeMsimp}).

We are interested in the case that the 5D Yukawa couplings are more or less universal
and the charge lepton mass hierarchy is achieved by different level of Gaussian overlap
between two chiral split fermions.
Therefore, we assume that the magnitudes of all the 5D Yukawa couplings are of order one.
We write  $\hat{f}_{ab}=\rho_{ab} e^{i\theta_{ab}}$, where $\rho_{ab}$ is a positive real random number in the
range of $[0.5,1.5]$. There is no reason of picking any specific complex phase for the Yukawa,
so we allow each  $\theta_{ab}$ to be randomly chosen from  $[0, 2\pi]$.

And we search for the configuration $\{ c^R_{1},c^R_{2}, c^R_{3},c^L_{1},c^L_{2},c^L_{3}  \}$, the central locations of
six split fermions,  and the 5D Higgs cubic coupling parameter $\kappa$.
For each set of configuration, we take a large number of random samples of 5D Yukawa couplings and check
whether the statistical outcome of charged lepton masses, neutrino masses, and neutrino mixings
all agree with the experimental data within 2 standard deviations.

We have found 4 such feasible configurations, see Table \ref{tab_4sets}.
Note that   in every configuration two left-handed fermions sit next to each other, within $0.5 \sigma$,
and the remaining one locates at a remote position from the two.
\begin{table}
  \centering
  \begin{tabular}{cccccccc}
Configuration  & $\kappa$ &  $c^R_{1}$ & $c^R_{2}$ & $c^R_{3}$ & $c^L_{1}$ & $c^L_{2}$ & $c^L_{3}$  \\
\hline\hline
I& $0.389$  & 10.112&2.989&9.592& 14.350 & 13.954 & 6.060 \\
\hline
II& $1.054$ & 9.789 & 9.570 & 10.557 & 5.715 & 13.498 & 5.201\\
 \hline
III& $0.169$ & 9.416& 8.956& 18.602& 5.881& 13.249& 13.591\\
 \hline
IV& $0.974$ & 1.371& 8.159& 17.663& 12.595& 12.106& 4.346\\
\hline
\end{tabular}
  \caption{The four viable configurations which can accommodate charge lepton and neutrino data in the same
  time. The split fermion location $c$'s are in the unit of $\sigma  ( = 5\times 10^{-4} R )$.}\label{tab_4sets}
\end{table}
The statistics of lepton masses and neutrino mixing angles are displayed in
Table \ref{tab:physobvs} and Table \ref{tab:NuMass}.
\begin{table}
  \centering
  \begin{tabular}{ccccccc}
Configuration & $m_e$(MeV) & $m_\mu$(MeV) & $m_\tau$(GeV) & $\sin^2 (2\theta_{12})$ & $\sin^2 (2\theta_{23})$ & $\theta_{13}$ ( rad)  \\
\hline\hline
I & $3.1\pm 1.5$ & $120(22)$ & $1.73(31)$& $0.79(24)$ & $0.43(26)$ & $0.11(8)$ \\
\hline
II& $6.3\pm 3.0$ & $119(20)$ & $2.49(48)$ & $0.84(18)$ & $0.72(24)$ & $0.16(11)$\\
 \hline
III& $0.64(12)$& $122(22)$ & $1.70(31)$& $0.76(27)$& $0.56(27)$& $0.33(20)$\\
 \hline
IV& $ 0.49(10)$& $78(14)$ & $2.25(43)$& $0.83(20)$ & $0.93(08)$& $0.13(7)$\\
\hline
\end{tabular}
  \caption{Charged lepton masses and neutrino mixings in the 4 viable configurations}\label{tab:physobvs}
\end{table}
The 5D scalar cubic coupling
$\kappa$ is found to vary from $\sim 0.1 - 1.0$ across the viable configurations.
Even for the case of $\kappa\sim 0.1$, one has the freedom to evenly distribute
the overall factor $(0.1)^{1/3}\sim 0.5$ to
$\kappa, \hat{f}^1, \hat{f}^h$ to avoid the order one fine turning.

In our numerical study, we only find neutrino masses of inverted hierarchy type, see Tab.\ref{tab:NuMass}.
The reason of getting the inverted hierarchy and the bi-large mixings in this model is different from  that of the
approximate $L_e-L_\mu-L_\tau$ symmetry in the Zee-Wolfenstein model.
The common feature in all four configurations
is the hierarchical 4D effective Yukawa couplings to $\Phi_2$ and $h$ in the charged lepton mass basis:
(1)$f^h_{e\mu} \gg f^h_{e\tau}, f^h_{\mu\tau}$, (2) $f^2_{\tau\tau} \gg $ all other $f^2$'s,
and (3) $f^h_{e\mu}\gg f^2_{\tau\tau}$.
This Yukawa couplings hierarchy results in the approximate neutrino mass matrix elements
\beqa
{\cal M}^\nu_{ee}\sim m_\mu f^2_{e\mu}f^h_{e\mu}\,,\;
{\cal M}^\nu_{\mu\mu}\sim m_\tau f^2_{\mu\tau}f^h_{\mu\tau}\,,\;
{\cal M}^\nu_{\tau\tau}\sim m_\mu f^2_{\tau\mu} f^h_{\mu\tau}\,,\nonr\\
{\cal M}^\nu_{e\mu}\sim m_\mu f^2_{\mu\mu}f^h_{e\mu}\,,\;
{\cal M}^\nu_{e\tau}\sim m_\mu f^2_{\tau\mu}f^h_{e\mu}\,,\;
{\cal M}^\nu_{\mu\tau}\sim ( m_e  f^2_{\tau e}f^h_{e\mu} + m_\tau f^2_{\tau\tau}f^h_{\tau\mu})\,,
\eeqa
where the common overall factors and the KK labels have been omitted.
From the above expression, we see that ${\cal M}^\nu_{ee}\sim  {\cal M}^\nu_{e\mu} \sim
{\cal M}^\nu_{e\tau} \gg  {\cal M}^\nu_{\mu\mu}, {\cal M}^\nu_{\tau\tau},{\cal M}^\nu_{\mu\tau}$.
The neutrino mass matrix is  Zee-Wolfenstein model like, except that  its  $ee$-component is  comparable to
the $e\mu$ and $e\tau$ components. The non-vanishing diagonal matrix elements are the key to accommodate the
bi-large mixings instead of the bi-maximal mixings, which is ruled out, in the Zee-Wolfenstein model.

Moreover, the first entry of neutrino Majorana mass matrix $|m^\nu_{ee}|$ is at the order of $0.01$ eV.
If we adopt a convention that all scalars carry no lepton number, only the interaction
$\bar{\Psi}^{\rm c} i\tau_2\Psi h$ breaks the lepton number by two units.
Since  $h^{(n)}$ is a singlet under $SU(2)_L$ and $SU(3)_c$, it has
no tree-level contribution to neutrinoless double beta decay ($0\nu\beta\beta$).
Therefore the ($0\nu\beta\beta$) process is dominated by the neutrino Majorana masses, namely, $|m^\nu_{ee}|$.
The effective neutrino mass  predicted by our model can be probed at the planned  detectors with few tons of isotope
( for references see \cite{0nuee}).

Since all the complex phases in our numerical are random, there is no prediction
for the three CP violation phases.

\begin{table}
  \centering
  \begin{tabular}{ccccc}
Configuration & $m^\nu_1$ & $m^\nu_2$ & $m^\nu_3$ &
$ |m^\nu_{ee}|$  \\
\hline\hline
I & $38\pm 13$ & $46\pm 14$& $1.4\pm 1.3$ & $14\pm 7$ \\
\hline
II & $41\pm 16$ & $45\pm 15$ & $5.1\pm 4.2$  & $6\pm 3$\\
 \hline
III & $40\pm 16$ & $45\pm 16$& $6.2\pm 5.0$ & $8\pm 4$\\
 \hline
IV & $39\pm 16$& $49\pm 15$& $5\pm 7$ & $9\pm 5$\\
\hline
\end{tabular}
  \caption{The absolute neutrino masses and the effective neutrino mass $|m^\nu_{ee}|$ (in  meV ).}\label{tab:NuMass}
\end{table}

\section{Phenomenology}

Once neutrinos are massive, the individual lepton number in SM is no longer conserved and
lepton flavor violating (LFV) is inevitable.
The current experimental upper limit on $Br(\mu\ra 3e)<10^{-12}$ and $Br(\tau\ra l_1 l_2 l_3) <3\times 10^{-8}$
\cite{Marciano:2008zz}
 set a stringent constraint on  new physics beyond SM.
\begin{figure}
\centering
\includegraphics[width=4.5in]{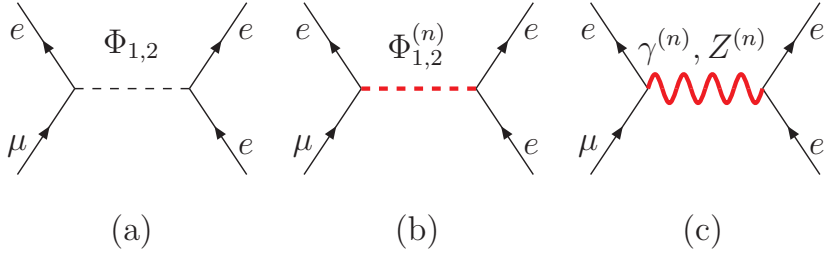}
\caption{Feynman diagrams for Lepton flavor violating $\mu\ra 3e$ decays in
(a) 4D Zee model (b)KK $\Phi_2$ mediated  (c)KK photon and  $Z$ mediated.
Red color (thick line) is used for the KK excitations. \label{fig:LFV}}
\end{figure}
In the 4D Zee model, the charged lepton masses receive contributions from both $v_1$ and $v_2$.
The mass matrix, ${\cal M}^e_{ab}=\frac{1}{\sqrt{2}}(y^1_{ab} v_1 +y^2_{ab} v_2)$,
can be diagonalized by some bi-unitary transformation:
$  V^\dag_L {\cal M}^e_{ab} V_R =\mbox{diag}\{m_e, m_\mu, m_\tau\}$.
Apparently, in the mass eigenbasis, both Yukawa $(V^\dag_L y^{1,2} V_R)_{ab}$  are not diagonal.
This is the root to the problematic FCNC in general 2HDM with electroweak Higgs, see Fig.\ref{fig:LFV}(a).

On the other hand, in the 5D Zee model  only the $Z_2$-even $\Phi^{(0)}_1$ can develop the VEV and take
the sole responsibility for the charged lepton masses and the electroweak symmetry breaking.
 In the mass basis, the fermion Yukawa couplings to the SM Higgs is always
diagonal.
However, the coupling of split fermion to the
KK gauge boson or scalar is proportional to the wave function overlap,
and the nontrivial profile of the KK mode will naturally generate the tree level FCNC couplings, Fig.\ref{fig:LFV}(b,c).
Therefore,  the  branching ratios of  the  tree-level LFV processes  will be much
larger than the loop induced radiative LFV transition,
e.g. $Br( \mu\ra 3e ) \gg Br(\mu\ra e \gamma)$.

In the mass basis, the gauge coupling to the KK gauge boson($n\geq 1$) can be approximated by
\beqa
\label{eq:LFVGC}
g^{(n)}_{L/R, ij}
&= & \sqrt{2}\,g^{SM}_{L/R} \sum_{a=1}^3 U_{L/R,ia}^\dag \cos\frac{n c^{L/R}_a}{R} U_{L/R,aj}\\
&\sim& \sqrt{2}\,g^{SM}_{L/R} \left[ \delta_{ij} -\frac{1}{2} \sum_{a=1}^3
U_{L/R,ia}^\dag \left( \frac{n c^{L/R}_a}{R}\right)^2 U_{L/R,aj}  +{\cal O}\left(\frac{\sigma^4}{R^4}\right)\right]\,,\nonr
\eeqa
where $g^{SM}_{L/R}$ is the  corresponding SM (chiral) gauge coupling constant.
A crucial difference between the effective gauge and Yukawa couplings
 is that the flavor diagonal effective 4D gauge coupling does NOT receive the exponential suppression since gauge interaction
couples exactly the same chiral fermion in the interaction basis. And the off-diagonal gauge coupling
starts at the first Taylor expansion of cosine function.
With these observations, we can estimate the relative size
 of LFV amplitudes mediated by KK gauge boson and KK scalars.
Taking $\mu\ra 3e$ as an example, the ratio of two amplitudes is estimated as
\beq
{{\cal M}^\gamma_{5D} \over  {\cal M}^{\Phi_2}_{5D}} \sim
{ e^2 \{ U^\dag_{L/R} [\cos\frac{\sigma}{R}] U_{L/R} \}_{\mu e} \over
 \frac{m_e}{v} \{ U^\dag_{L/R} [\sin\frac{\sigma}{R}] U_{R/L} \}_{\mu e} }
 \sim {4\pi\alpha \over (m_e m_\mu /v^2)} {(\sigma/R)^2 \over (\sigma/R) }\sim  10^4\,,
\eeq
where we take $e$ for the flavor diagonal KK gauge photon coupling,  $y_{ee}\sim m_e/v$
for the flavor diagonal KK Yukawa coupling,  $m_\mu/v$ for the Gaussian suppression between two
fermions with different chiralities, and Taylor expansion  $(\sigma/R)^{1,2}$ for sine/cosine function
for the flavor changing Yukawa/gauge couplings. Therefore, in the 5D Zee model the KK photon or KK $Z^0$ gauge
boson  mediated FCNC processes  are the most
dominate one.
Now we compare the  FCNC mediated  by the KK gauge bosons in the 5D Zee model to the
  FCNC mediated by the electroweak Higgs in the 4D Zee model.
The ratios of the two amplitudes can be estimated by the same token to be:
\beqa
{{\cal M}^\gamma_{5D} \over  {\cal M}^{H}_{4D}} &\sim&
{ e^2  \{ U^\dag_{L/R} [\cos\frac{\sigma}{R}] U_{L/R} \}_{\mu e} /(1/R)^2 \over
 \{ U^\dag_{L/R} [f] U_{R/L} \}_{\mu e} (m_e/v)/M_H^2}
\sim { 4\pi \alpha  \{ U^\dag_{L/R} \frac{\sigma^2}{R^2} U_{L/R} \}_{\mu e} /(1/R)^2 \over
 \{ U^\dag_{L/R} \frac{ m_\mu}{v} U_{R/L} \}_{\mu e} (m_e/v) /M_H^2}\nonr\\
& \sim & 4\pi\alpha {  M_H^2 \over (1/R)^2 } {(\sigma/R)^2 \over (m_e m_\mu/v^2) }\sim 10^{-1}\,.
\eeqa

Therefore, in terms of the LFV branching ratio, the 5D Zee model is $\sim 10^{-2}$ smaller
than the 4D Zee model. Comparing to its 4D ancestor, this suppression gives us more allowance
to make the model phenomenologically viable.
Following the standard notation \cite{Chang:2005ag},
LFV branching ratios can be easily calculated
by summing up the contributions from the first $200$ KK photon and $Z$ excitations.
The result is summarized in Tab.\ref{tab_LFV}.
We do not include the rare tau decays to $\tau \ra e^-\mu^+ e^-, e^-e^+\mu^-, \mu^-e^+\mu^-, \mu^-\mu^+e^-$
for they are doubly suppressed by two LFV vertices.
From Fig.\ref{fig:LFV}(c), we expect that  $Br(\tau\ra \mu^- \mu^+\mu^-) \sim Br(\tau\ra \mu^- e^+e^-)$
and $Br(\tau\ra e^- \mu^+\mu^-) \sim Br(\tau\ra e^- e^+e^-)$, and the expectation is supported by the numerical, within factor
3.
However, there is no obvious pattern among $Br(\mu\ra e^- e^+e^-)$, $Br(\tau\ra \mu^- \mu^+\mu^-)$,  and $Br(\tau\ra e^- e^+e^-)$
 across the four configurations we found.
 Instead, the relative sizes of different LFV branching ratios can provide
a handle to distinguish different geography in the split fermions scenario.
This shows the importance of improving of the current LFV experimental bounds. They will
provide crucial information  to decipher the origin of flavor physics.
\begin{table}
  \centering
  \begin{tabular}{lcccc}
\hline\hline
Decay mode & Conf. I  & Conf. II & Conf. III & Conf. IV  \\
\hline\hline
 $Br(\mu^- \rightarrow e^+e^-e^-)$& $4 (2) \times 10^{-13}$ & $1.6(6) \times 10^{-13}$ & $2(1) \times 10^{-13}$ & $1.3(7) \times 10^{-13}$  \\
 \hline
  $Br(\tau^-\rightarrow e^+ e^- e^- )$& $1.9(9) \times 10^{-11}$ & $9(6) \times 10^{-14}$ & $1.5(1.5) \times 10^{-14}$ & $1.3(1.3) \times 10^{-18}$  \\
 \hline
 $Br(\tau^- \rightarrow \mu^+\mu^- e^-)$& $1.0(5)\times 10^{-11}$ & $5(3) \times 10^{-14}$ & $1.0(9) \times 10^{-14}$ & $1.2(1.2) \times 10^{-18}$  \\
\hline
$Br(\tau^- \rightarrow e^+e^- \mu^-)$& $4(3) \times 10^{-13}$ & $3.0(2.8) \times 10^{-14}$ & $2.8(2.6) \times 10^{-13}$ & $3(2) \times 10^{-13}$  \\
\hline
 $Br(\tau^-\rightarrow\mu^+ \mu^- \mu^- )$&$7(6) \times 10^{-13}$ & $5.3(5.0) \times 10^{-14}$ & $7(6) \times 10^{-13}$ & $1.1(6) \times 10^{-12}$  \\
 \hline
\end{tabular}

  \caption{Lepton flavor violating decays}\label{tab_LFV}
\end{table}

Now we change gear to collider signals.
There is only one SM like neutral Higgs  below or at the electroweak scale due to
the specific orbifold boundary condition we chose.
Around ${\cal O}(1/R)\sim $ TeV, there are 10 physical heavy Higgs (the first KK excitations
of three kinds of scalars) accompanying  the appearance of four KK gauge bosons, $\gamma^{(1)}, Z^{(1)}$,
and $W^{\pm,(1)}$ as in any extra dimension model. The  study of these KK excitations
at collider depends on the model details and it is beyond the scope of this letter.
Notice that  replacing the electron current at the right hand side of Fig.\ref{fig:LFV}(c)
by the initial state of $q\bar{q}$ (for the LHC) or  $e\bar{e}$ (for the linear collider )
leads to 1st KK gauge boson LFV decays.
The  branching ratio of LFV decay $V^{(1)}\ra l^+_i l^-_j$ is proportional to
$\left|g^{V_1}_{L,ij}\right|^2+ \left|g^{V_1}_{R,ij}\right|^2$.
The KK gauge couplings $g^{V_1}_{L,ij}$ and $g^{V_1}_{R,ij}$, given in Eq.(\ref{eq:LFVGC}),
depend on the detailed  split fermion configuration.
However,  because $\sin^2\theta_W=0.23 \sim 1/4$, so $\left|-1/2 +\sin^2\theta_W\right| \sim \left|\sin^2\theta_W\right|$,
the  LFV branching ratios of the first  KK photon and the first  KK Z are proportional to each
other mode by mode.
Moreover, in the case that LFV is dominated by the first KK gauge boson,
there is one interesting connection between the LFV decays of the first KK gauge boson $V^{(1)}$, be it the $\gamma^{(1)}$ or $Z^{(1)}$,
and  in the rare tau/muon decay:
 \beq
 { Br(V^{(1)} \ra \tau e)\over Br(V^{(1)} \ra \mu e) } : { Br(V^{(1)} \ra \tau \mu) \over Br(V^{(1)} \ra \mu e) }
 \sim  { Br(\tau\ra 3e)\over  Br(\mu\ra 3e)}:{Br(\tau\ra 3\mu) \over  Br(\mu\ra 3e)}\,.
 \eeq
The reason is clear by looking at Fig.\ref{fig:LFV}(c).

Taking the acceptance of CMS and ATLAS into account, it is estimated to have
roughly $\sim 4\times 10^{4}\times (\sqrt{2})^4=1.6\times 10^5 $
expected events of 1TeV KK  boson to be observed at the LHC with $\sqrt{s}=14 TeV$ and $100 fb^{-1}$ luminosity\cite{Dittmar:2003ir}
\footnote{The enhancing factor $(\sqrt{2})^4$ is due to the couplings of KK gauge boson.}.
Unfortunately, to satisfy the stringent bounds on muon/tau LFV decays,
 the LFV branching ratios  of $\gamma^{(1)}, Z^{(1)}$
 are found to be smaller then $10^{-12}$   for all the 4 configurations.
Which basically makes the observation
of LFV decay $V^{(1)}\ra e\mu, \mu\tau, e\tau$ impossible at the LHC or
the planned linear collider.

\section{Conclusion and discussion}
The Zee model is a very economical extension of the SM to generate neutrino mass without
the presence of right-handed neutrinos.
However, the original model  in the 4D spacetime faces two major difficulties: the tree-level FCNC mediated by
electroweak Higgs and too many arbitrary Yukawa.
In this work, we study the Zee model in a 5D split fermions scenario
where each SM chiral fermion has a Gaussian profile
localized in the $S_1/Z_2$ orbifold, with a compactification radius $R$ .
For simplicity, all split fermions profile in 5D are assumed to have a common width $\sigma$.
The SM $SU(2)_L$ and $U(1)_Y$ gauge fields and all the three Higgs fields in the Zee model are
now propagating in the fifth dimension.
The second Higgs doublet and the Zee singlet are assigned to be odd under $Z_2$ orbifolding, such that
their zero modes are projected out. Only the first Higgs doublet ( whose zero mode is identified as the SM Higgs)
can develop VEV to give charged fermion masses and break the SM electroweak symmetry.

We assume that all the 5D Yukawa couplings are complex number
with arbitrary phase and their absolute values of order one.
With only 9 important  parameters ( $1/R, \sigma/R, \kappa$ and six split fermion locations), we are able to accommodate
the three charged lepton masses and all observed neutrino data simultaneously
without much fine tuning.
This largely reduces the number of unknown parameters
(  mostly Yukawa ) in the 4D version.
It is highly nontrivial that such kind of solutions exist in this model. For instance, when the Zee model is embedded
in an $SU(3)_W$ GUT model\cite{Chang:2003sx}, no viable solution in the split fermion scenario has been found.

Whether the neutrino mass of normal hierarchy type can be accommodated in this model is not clear to us.
However, only the solutions for  neutrino mass of inverted hierarchy type were found in our numerical study
so far. In all of the four representative configurations,  nonzero $\theta_{13}$ and
the effective neutrino mass $|m_{ee}|\sim 0.01$eV
were predicted.  Which  may be probed at the planned reactor neutrino experiments and the
$(0\nu\beta\beta)$  experiments with few tons of  isotope.

Below the electroweak scale, the SM Higgs coupling is flavor diagonal thus no FCNC processes  at tree level.
In this model, the  LFV processes are dominated by the tree level diagrams  mediated by first KK gauge boson $V^{(1)}$ ( be it the photon or the $Z$ ),
which are suppressed compared to  the 4D version due to (1) the heavy mass ( $\sim$ TeV )
and (2) the narrow width of the Gaussian distribution of split fermions. We find no pattern in the LFV
rare tau or muon decay channels. Instead, once been measured, the relative magnitude amount branching ratios
can be used to tell which configuration is preferred by the nature.

The direct signal of extra dimension will be the discovery of the KK excitations of gauge bosons or scalars
at the high energy colliders.  However, in the foreseeable future, it is more promising to study the lepton flavor physics
in the LFV rare tau/muon decays  for this model.

In all, this model provides a natural and economical setup for generating neutrino masses through the Zee mechanism
with
(1) a small number of free parameters,
(2) the hierarchical 4D Yukawa couplings,
(3) the absence of the tree-level FCNC at low energy, $\leq v$, and the naturally heavy exotic particles.
It also demonstrates a  geometrical alternative to accommodate the observed pattern
in lepton masses and mixings without any flavor symmetry involved.
Although the split fermion model with a universal Gaussian width is rather ad hoc, we believe the idea presented in this work
can be applied to other higher-dimensional models without major difficulties.

\begin{acknowledgments}
Work was supported in part by the Taiwan NSC under Grant No.\
99-2112-M-007-006-MY3.
\end{acknowledgments}

\end{document}